\documentclass[onecolumn,aps,superscriptaddress,prd,10pt,showpacs,preprintnumbers,nofootinbib,eqsecnum]{revtex4-2}
\usepackage{graphicx,slashed,bbold,subfigure,amsmath,amssymb,enumerate,color,xcolor}
\usepackage{multirow}
\usepackage{natbib,ulem} 
\usepackage[colorlinks=true,citecolor=blue,linkcolor=blue,urlcolor=blue]{hyperref}
\setcitestyle{square, numbers, comma, sort&compress}

\begin{document}
  
\title{Non-strange quark stars within resummed QCD} 
\author{Tulio E. Restrepo}
 \email{tulioeduardo@pos.if.ufrj.br}
\affiliation{Instituto de F\'isica, Universidade Federal do Rio de Janeiro, Caixa Postal 68528, 21941-972, Rio de Janeiro, RJ, Brazil}

\author{Constan\c{c}a Provid\^{e}ncia}%
 \email{cp@uc.pt}
 \affiliation{CFisUC, Department of Physics, University of Coimbra, 3004-516 Coimbra, Portugal}
 
 \author{Marcus Benghi Pinto}
\email{ marcus.benghi@ufsc.br}
\affiliation{Departamento de F\'isica, Universidade Federal de Santa Catarina, 88040-900 Florian\'opolis, SC, Brazil}

\begin{abstract}
The recently developed resummation technique known as  {\it renormalization group optimized perturbation theory} (RGOPT)  is employed in the evaluation of the EoS describing non-strange cold quark matter at NLO. Inspired by recent investigations, which suggest that stable quark matter can be made only of up and down quarks, the mass-radius relation for two flavor pure quark stars is evaluated and compared with  the predictions from perturbative QCD (pQCD) at NNLO. This comparison explicitly shows that  by being imbued with renormalization group properties, and  a variational optimization procedure, the method  allows for an efficient resummation of the perturbative series. Remarkably, when the renormalization scale is chosen so as to reproduce  maximum mass stars with M=$2-2.3M_\odot$, one obtains a mass-radius curve compatible with the  masses and radii of the pulsars PSR J0740+6620, PSR J0030+0451, and the  compact object HESS J1731-347. Moreover, the scale dependence  of the EoS (and  mass-radius relation) obtained with the RGOPT is greatly improved when compared to that of pQCD.  This seminal application to the description of quark stars shows that the RGOPT represents a robust alternative to pQCD when describing compressed quark matter.
\end{abstract}

\maketitle
\section{Introduction}
The knowledge of the equation of state (EoS) for cold ($T=0$) and dense quark matter is necessary for the description of compact stellar objects (CSO) such as neutron and quark stars. 
The recent observational data of nearly two solar mass pulsars \cite{Demorest:2010bx,Fonseca:2016tux,Antoniadis:2013pzd,Romani:2021xmb} suggests that {quite high baryonic densities may be attained inside neutron stars, and, in particular, that  quark matter may be present inside CSO.}
Furthermore, the detection of gravitational waves (GW) from binary neutron star mergers by the LIGO-Virgo Collaboration (LVC) \cite{Abbott:2018exr,LIGOScientific:2018hze}, together with NICER x-ray determination of the mass and radius of two neutron stars \cite{Nattila:2017wtj,Riley_2019,Miller:2019cac,Riley:2021pdl,Miller:2021qha} give important constraints on the possible EoS for quark and neutron stars \cite{Annala:2017llu,Radice:2017lry,Dietrich:2020efo,Landry:2018prl,Annala:2021gom}.

Since the Bodmer--Witten hypothesis, that stable quark matter (QM) must contain strange quark matter (SQM) \cite{Bodmer:1971we,Teraza1979,Witten:1984rs}, most authors have described  quark stars (QS) and neutron stars (NS) as containing SQM,  see for instance \cite{Glendenning:2000}.
The Bodmer--Witten hypothesis states that the energy per baryon at zero pressure must be lower than the one representing the most stable nucleus $^{56}{\rm Fe}$, $\varepsilon(P=0)=E/A\leq 930$ MeV.
However, a recent investigation opens the possibility that $ud$ quark matter, or non-strange QM (NSQM), can be the ground state of cold and dense baryonic matter at zero pressure \cite{Holdom:2017gdc}. In this vein, the authors of Ref. \cite{Holdom:2017gdc} use a phenomenological  model to discuss the QCD spectrum and the conditions for NSQM to be the ground state, $\varepsilon_{\rm NSQM}\leq 930 \ {\rm MeV}\leq \varepsilon_{\rm SQM}$.
More recently, the properties of non-strange QS (NSQS) have also been explored by other authors \cite{Zhao:2019xqy,Wang:2019jze,Zhang:2019mqb,Ren:2020tll,Wang:2020wzs,Xu:2021alh,Zhang:2020jmb,Yuan:2022dxb,Cao:2020zxi,Xia:2022tvx}. 

From the theoretical point of view, the description of compact stellar objects is commonly described by chiral effective models, in the low density regime, \cite{Epelbaum:2008ga,Machleidt:2011zz} and by perturbative QCD (pQCD) at very high densities \cite{Kraemmer:2003gd}. 
As an alternative to these two approaches, one can resort to effective models such as the MIT bag model \cite{Chodos:1974pn}, the Nambu--Jona-Lasinio (NJL) model \cite{Nambu:1961tp,*Nambu:1961fr}, as well as the linear sigma model \cite {Gell-Mann:1960mvl}, among others. 
As far as QCD is concerned, a seminal work  by Freedman and McLerran has provided the  next-to-next-to leading order (NNLO) pressure for (massless) cold quark matter \cite{Freedman:1976dm,Freedman:1976ub}. The results have been later refined so as to include a massive strange quark as well as massless $ud$ quarks  in a $N_f=2+1$ application \cite{Fraga:2004gz,Kurkela:2009gj,Fraga:2013qra}. 
After more than four decades, coefficients for the $\text{N}^3$LO contribution to the pressure of massless quarks have been recently calculated \cite{Gorda:2018gpy,Gorda:2021znl,Gorda:2021kme}. At the same time, the leading and next-to-leading logarithmic soft contributions have been resummed to all orders in Ref. \cite{Fernandez:2021jfr}, improving the pQCD state of the art result from Ref. \cite{Gorda:2021znl,Gorda:2021kme}.

For many years, the community that studies CSO has adopted the MIT model to describe QS, hybrid stars and also the collapse of supernovae nucleus \cite{Witten:1984rs,Alcock:1986hz,Haensel1986,Berezhiani:2002ks,Alford:2004pf,Drago:2005yj,Mintz:2009ay}. 
Unfortunately, in its pure form this model predicts QS masses below $2M_\odot$. 
On the other hand, when the $\overline{\rm MS}$ renormalization scale ($\Lambda$) is taken at the so-called ``central" value, $(2/3)(\mu_u +\mu_d +\mu_s)$, the NLO pQCD EoS predicts QS masses above $2M_\odot$ \cite{Fraga2001,Fraga:2001xc,Fraga:2004gz} and above $2.5M_\odot$ at NNLO \cite{Kurkela:2009gj,Jimenez:2019iuc,Jimenez:2021wik}. 
However, the downside of pQCD it that its results are highly dependent on the renormalization scale value, which in general is taken to lie between half and twice the central value. 

Few years ago, the {\it optimized perturbation theory} (OPT)  \cite{Okopinska:1987hp,Duncan:1988hw} was substantially improved in order to include renormalization group (RG) properties \cite{Kneur:2010ss}. The resulting approximation has been dubbed {\it renormalization group optimized perturbation theory} (RGOPT).
Within QCD, at $T=\mu=0$, the RGOPT was first used to calculate with excellent certainty the QCD scale, $\Lambda_{\overline{\text{MS}}}$, in the $\overline{\text{MS}}$-scheme \cite{Kneur:2011vi}. 
Subsequently, the strong coupling constant, $\alpha_s$, was calculated from the pion decay constant \cite{Kneur:2013coa}, $F_\pi$, while very accurate results for the  quark condensate were generated from the evaluation of the spectral function, up to six loop order \cite{Kneur:2015dda,Kneur:2020bph}. 
In later applications, the RGOPT was applied to scalar field theories at finite temperature \cite{Kneur:2015moa,Kneur:2015uha,Ferrari:2017pzt,Fernandez:2021sgr}, showing explicitly how the method highly improves the scale dependence in thermal theories when compared to {\it screened perturbation theory} (SPT) \cite{Karsch1997,Andersen:2000yj} and  {\it hard thermal loop perturbation theory} (HTLpt) \cite{Pisarski:1988vd,Braaten:1989kk,Braaten:1989mz,Braaten:1990az}.
The method was also employed to describe  cold and dense QCD  \cite{Kneur:2019tao}, as well as the quark sector of QCD at finite temperatures and baryonic densities \cite{Kneur:2021dfo,Kneur:2021feo}, producing results which are undoubtedly less scale dependent than those furnished by pQCD and HTLpt . 

In this work, our main goal is  to use the RGOPT to evaluate the QCD EoS  in order to describe NSQS. Aiming to explicitly display the advantages of the method, we confront the RGOPT predictions, at zero temperature and finite densities, with those furnished by pQCD and compare with the most recent CSO observational data. 
 Following previous pQCD applications \cite{Fraga:2004gz,Kurkela:2009gj,Fraga:2013qra} to the case $N_f=2+1$ we consider the $ud$ quarks to be massless in our present $N_f=2$ application. In this situation the RGOPT results  of Ref. \cite{Kneur:2019tao}, originally obtained for $N_f=3$, can  be trivially adapted to the $N_f=2$ case at hand. 
As it will be shown, when the renormalization scale is appropriately chosen the RGOPT resummation of the perturbative series is able to describe several  observational data, in particular from  pulsars  PSR J0030+0451 \cite{Riley_2019,Miller:2019cac}, PSR J0740+6620 \cite{Cromartie:2019kug,Riley:2021pdl,Miller:2021qha} and the light compact object recently identified in the middle of a supernova remnant,  HESS J1731-347 \cite{2022NatAs.tmp..224D}.   Notice, however, that this last star is supposed to be a hot star and the direct Urca processes in a QS is very efficient to cool down the star, unless quarks are paired (see also \cite{DiClemente:2022wqp} where the authors advocate that  HESS J1731-347 could be a strange quark star). 
Finally, keeping  in mind that the renormalization scale is chosen to be  density dependent we also discuss how  the EoS, for both RGOPT and pQCD, has to be handled   so that thermodynamic consistency is preserved. 

The work is organized as follows. In Sec. \ref{RGOPTpres} we review the basics of the RGOPT implementation in the context of cold and dense quark matter. In Sec. \ref{therconsis}, a thermodynamic consistent EoS is constructed for both approximations. In Sec. \ref{results}, we present our numerical results discussing the thermodynamic consistent EoS and the associated  mass-radius relations. Finally, in Sec. \ref{conclusion}, we draw our conclusions and present some perspectives.
\section{RGOPT quark pressure}\label{RGOPTpres}

In order to make the manuscript self-consistent let us start by reviewing some of the main results contained in Ref. \cite{Kneur:2019tao}.
Before applying the RGOPT prescription, one must know the {\it massive} perturbative expression for the physical quantity one wants to optimize, even when the theory to be resummed contains massless particles . 
With this aim one may start by considering   the perturbative  massive pressure {\it per flavor} at order $\alpha_s$ and finite chemical potential. Considering the degenerate two flavor case, $f=u,d$ and $m_u=m_d \equiv m$, the pressure is easily obtained by combining the vacuum ($\mu_f=0$) results from Ref. \cite{Kneur:2015dda} and the in-medium ($\mu_f\ne 0$) results from Refs.\cite{Akhiezer:444287,Farhi:1984qu,Fraga:2004gz} 
 \begin{align}
\begin{split}
 P^{\rm{PT}}_{1,f}(\mu_f)=&-N_c \frac{m^4}{8 \pi^2} \left(\frac{3}{4}-L_m\right)+\Theta(\mu_f^2-m^2)\,\frac{N_c}{12 \pi^2} 
\left [ \mu_f p_F\left (\mu_f^2 - \frac{5}{2} m^2 \right ) + \frac{3}{2} m^4 \ln ( \frac{\mu_f +p_F}{m} ) 
\right ] \\
 &- \frac{d_A \,g}{4\left(2\pi\right)^4} m^4 \left(3 L_m^2-4 L_m+\frac{9}{4}\right) 
 -\Theta(\mu_f^2-m^2)\, \frac{d_A\, g}{4\left(2\pi\right)^4} \left \{ 3 \left [m^2 \ln ( \frac{\mu_f +p_F}{m} ) 
 -\mu_f p_F\right ]^2 - 2 p_F^4 \right \}
  \\
&- \Theta(\mu_f^2-m^2)\, \frac{d_A\,g}{4\left(2\pi\right)^4} m^2  \left ( 4- 6 L_m \right ) 
 \left [ \mu_f p_F - m^2 \ln ( \frac{\mu_f +p_F}{m} ) \right ] .
\label{PPTqcd} 
\end{split}
\end{align}
where $p_F=\sqrt{\mu_f^2-m^2}$ is the Fermi momentum, $L_m=\ln(m/\Lambda)$, $d_A=N_c^2-1$, while $\Theta$ represents the Heaviside function and $g=4\pi\alpha_s$ is our definition for the coupling\footnote{Even though most  works use the notation $g_s^2=4\pi\alpha_s$, we adopt the convention $g=g^2_s$ instead in order to be consistent with the notation used in previous RGOPT applications to QCD. }.
For $\mu_f>m$, this expression can be further simplified using the properties of logarithmic function: 
\begin{align}
\begin{split}
 P^{\text{PT}}_{1,f}(\mu_f)=& 
  \frac{N_c}{12 \pi^2} \left [ \mu_f p_F\left (\mu_f^2 - \frac{5}{2} m^2 \right ) 
  + \frac{3}{2} m^4 \left ( L_\mu -\frac{3}{4} \right ) \right ]\\
  &-\frac{g d_A}{4\left(2\pi\right)^4}\left[ m^4 \left ( 3 L_\mu^2- 4 L_\mu + \frac{1}{4} \right ) +
 \mu_f^2 \left(\mu_f^2+m^2 \right) +m^2\,\mu_f \, p_F \left(4 - 6 L_\mu \right) \right]  ,
\end{split}
 \end{align}
where $L_\mu \equiv \ln [(\mu_f+p_F)/\Lambda]$.

As explained in Ref. \cite{Kneur:2019tao} the observable represented by the perturbative pressure, Eq. (\ref{PPTqcd}), is not invariant under the action of  the RG operator
\begin{align}
 \Lambda \frac{d}{d \Lambda}= \Lambda \frac{\partial}{\partial \Lambda} + \beta(g )\frac{\partial}{\partial g } - 
\gamma_m(g ) m \frac{\partial}{\partial m} \;,
\label{RG}
\end{align}
in contradiction with what is expected from RG properties.
Then, to obtain a RG invariant (RGI) pressure, one must add zero point contributions to the quark pressure \cite{Kneur:2015dda,Kneur:2015moa,Kneur:2015uha}, such that
\begin{align}
 P^{\rm{PT}}_{1,f}(\mu_f) \to P^{\rm{RGI}}_{1,f}(\mu_f)=P^{\rm{PT}}_{1,f}(\mu_f) -m^4 \sum_k s_k g^{k-1}\; 
 \label{sub}
\end{align}
is completely scale invariant. As emphasized in Refs. \cite{Kneur:2015dda,Kneur:2015moa,Kneur:2015uha}, the contribution given by the vacuum terms in Eq. (\ref {PPTqcd}) are of utmost importance to determine the coefficients $s_k$, which depend on RG coefficients of the $\beta$ and $\gamma_m$ RG functions.
 In our notation the $\beta$ and $\gamma_m$ functions are defined by
\begin{equation}
 \beta\left(g\equiv 4\pi\alpha_s \right)=-2b_0g^2-2b_1g^3+\mathcal{O}\left(g^4\right) \;,
 \label{betaQCD}
 \end{equation}
 and
 \begin{equation}
 \gamma_m\left(g\right)=\gamma_0g+\gamma_1g^2+\mathcal{O}\left(g^3\right)\;,
\end{equation}
where the coefficients read
\begin{align}
 b_0=& \frac{1}{\left(4\pi\right)^2}\left(11-\frac{2}{3}N_f\right ),\\
 b_1=& \frac{1}{\left(4\pi\right)^4}\left (102-\frac{38}{3}N_f\right ),\\
 \gamma_0=&\frac{1}{2\pi^2},
 \end{align}
 and
\begin{equation}
 \gamma_1^{\overline {\rm MS}}= \frac{1}{8\left(2\pi\right)^4}\left (\frac{202}{3}-\frac{20}{9}N_f\right) \;.
\end{equation}
The coefficients $s_k$ are found by applying the RG operator, Eq. (\ref{RG}), to Eq. (\ref{sub}) such that it vanishes up to  $\mathcal{O}(g^2)$. This procedure yields
\begin{equation}
 s_0=- N_c \left[(4\pi)^2 (b_0-2\gamma_0)\right]^{-1} \;,
 \end{equation}
 and
 \begin{equation}
 s_1=-\frac{N_c}{4}\left[\frac{b_1-2\gamma_1}{4(b_0-2\gamma_0)} -\frac{1}{12\pi^2}\right ]\,.
 \label{s1}
\end{equation}
Once the RGI pressure is obtained, one proceeds by  adding a modified Gaussian interpolating  term so that one has
\begin{align}
 {\cal L}_{\rm{QCD}}^{\rm{RGOPT}}= {\cal L}_{\rm{QCD}}\vert_{g \to \delta g}^{m\to m+\eta\left(1-\delta\right)^a},
 \label{Lint}
\end{align}
where $\delta$ is a dummy parameter (taken to be small and later set to the unit value) while $\eta$ represents an {\it arbitrary} mass parameter which is fixed by a variational criterion\footnote{Here we adopt a notation slightly different from the one employed in  previous RGOPT applications where the variational mass parameter was denoted by $m$. In order to avoid confusion between the RGOPT variational mass (a Lagrange multiplier) and the original current quark masses (a physical quantity), we employ the notation used in many OPT works denoting the former by $\eta$ and the latter by $m$.}. At the same time the exponent $a$ is variationally fixed to $a= \gamma_0/(2b_0)$ (see Refs. \cite{Kneur:2015dda,Kneur:2015moa,Kneur:2015uha,Kneur:2019tao} for further details).
In practice, the RGOPT prescription is applied by making the replacements, $m\to m+\eta\left(1-\delta\right)^a$  and $g\to \delta g$ in the RGI massive pressure (\ref{sub}). After that one can safely consider the chiral limit by setting $m=0$ since now $\eta$ naturally regularizes the quark propagator in the infra-red limit. Then, any physical quantity evaluated with the resulting  theory can be expanded to the desired order in powers of  the bookeeping parameter, $\delta$.
By applying the RGOPT prescription to Eq. (\ref{sub}) one obtains \cite{Kneur:2019tao}
\begin{align}
\begin{split}
 P^{\rm{RGOPT}}_{1,f}(\mu_f)=& P^{\rm{RGOPT}}_{0,f}(\mu_f) 
- N_c \frac{\eta^4}{\left(4\pi\right)^2}  \left (\frac{\gamma_0}{b_0} \right ) \left ( \frac{1}{b_0\,g} \right)
-N_c \frac{\eta^4}{4} \left (2\frac{\gamma_0}{b_0} -1\right ) \left(\frac{b_1-2\gamma_1}{4(b_0-2\gamma_0)} -\frac{1}{12\pi^2}\right ) \\
&+ N_c \frac{\eta^2}{8\pi^2} \left( \frac{\gamma_0}{b_0}\right ) 
\left [ \eta^2 \left( 1 - 2L_\mu \right ) +2 \mu_f \, p_F  \right ]  \\ 
 &- \frac{g d_A} {4\left(2\pi\right)^4} \left[ \eta^4 \left (\frac{1}{4} - 4 L_\mu +3 L_\mu^2 \right ) +
 \mu_f^2 \left(\mu_f^2+\eta^2 \right) +\eta^2\,\mu_f \, p_F \left(4 - 6 L_\mu \right) \right]   \;,
 \end{split} 
\label{rgoptNLO}
\end{align}
where 
\begin{equation}
  P^{\rm{RGOPT}}_{0,f}(\mu_f)=
  N_c \frac{\eta^4}{\left(4\pi\right)^2 b_0 \, g} + 
  \frac{N_c}{12 \pi^2} \left [ \mu_f p_F\left (\mu_f^2 - \frac{5}{2} \eta^2 \right ) 
  + \frac{3}{2} \eta^4 \left ( L_\mu -\frac{3}{4} \right ) \right ]\;,
 \label{rgoptLO} 
\end{equation}
is the LO expression for the RGOPT pressure. Note that now the Fermi momentum appearing in Eqs. (\ref{rgoptNLO}) and (\ref {rgoptLO}) reads $p_F=\sqrt{\mu_f^2-\eta^2}$.
\subsection{Optimization}
At NLO, the optimized variational mass, $\bar\eta$, is generally fixed by solving the relation 
 \begin{equation}
f_{\rm{MOP}} = \frac{\partial {P}^{\rm{RGOPT}}}{\partial \eta} \equiv 0 \;,
\label{mop}
\end{equation} 
which is known as the mass optimization prescription (MOP).
At the same time, the value of the optimized coupling constant, $\bar g$, is the solution of the {\it reduced} RG equation,
\begin{equation}
f_{\rm{RG}} \equiv \Lambda \frac{\partial  P^{\rm{RGOPT}} }{\partial \Lambda} + \beta(g)\frac{\partial 
P^{\rm{RGOPT}}}{\partial g } \equiv 0 \;.
\label{red}
\end{equation}
Unfortunately,  when going beyond LO one hits spurious imaginary solutions in the whole relevant $\mu$ range, as explained in Ref. \cite{Kneur:2019tao}. 
This problem, which is not exclusive of the RGOPT, is also usually found in other resummation techniques such as HTLpt and SPT. In order to find real solutions, one alternative is to execute a renormalization scheme change (RSC), which consists in shifting the mass parameter, $m$, in the massive RGI pressure, Eq. (\ref{sub}). In this case, following  Ref. \cite{Kneur:2013coa}, one performs the shift
\begin{align}
 m\to m (1+ B_2 g^2)\,,
 \label{RSC}
\end{align}
after which  the series is re-expanded up to order-$g$. This procedure implies the addition of an extra term,   $- 4g m^4 s_0 B_2 $, to the pressure given by Eq. (\ref{sub}) or, equivalently, a $- 4g \eta^4 s_0 B_2$ term to the RGOPT pressure, Eq. (\ref{rgoptNLO}). 
The new parameter $B_2$ is in principle arbitrary, but it must be sufficiently small to assure  that the RSC is perturbative. One way to fix $B_2$ is to require the contact of $f_{\rm {MOP}}$ and $f_{\rm{RG}}$ 
\begin{equation}
f_{\rm{RSC}}= \frac{\partial f_{\rm{RG}}}{\partial g} \frac{\partial f_{\rm{MOP}}}{\partial \eta} - 
\frac{\partial f_{\rm{RG}}}{\partial \eta} \frac{\partial f_{\rm{MOP}}}{\partial g} \equiv 0\,.
\label{contact}
\end{equation}
However,  once again the simultaneous evaluation of Eqs. (\ref{mop}), (\ref{red}) and (\ref{contact}) lead to imaginary solutions in the relevant density range. In this situation, one can  adopt the strategy employed in  Ref. \cite{Kneur:2019tao} where only Eqs. (\ref{mop}) and (\ref{contact}) are considered in order to optimize $\eta$ and $B_2$. When this strategy is adopted, the coupling, $g$, is simply given by its two loop perturbative expression
\begin{align}
 \ln \frac{\Lambda}{ \Lambda_{\overline{\text{MS}}} } = \frac{1}{2b_0\, g} +
\frac{b_1}{2b_0^2} \ln \left ( \frac{b_0 g} {1+\frac{b_1}{b_0} g} \right) .
\label{g2L}
\end{align}
Here we choose the value $\alpha_s(\Lambda_0=1500 \ {\rm MeV})=0.326$ \cite{Bazavov:2012ka}, so that $\Lambda_{\overline{\text{MS}}}\simeq 389\,{\rm MeV} $ for $N_f=2$. After these successive approximations one should expect that some residual scale dependence will show up in the final results. Nevertheless, as we shall explicitly demonstrate, the RGOPT remnant scale dependence is milder than the one plaguing the pQCD results.
\section{Thermodynamic consistency} \label{therconsis}

Before considering the NLO RGOPT EoS to evaluate the mass-radius relation let us recall that when using approximations like pQCD, HTLpt and RGOPT one usually chooses $\Lambda$ to depend on control parameters such as $T$ and $\mu$. However, this procedure may spoil the thermodynamic consistency of the EoS  since the thermodynamic relation for the energy density \cite{landau1980statistical},
\begin{align}
 \mathcal{E}=T\frac{dP}{dT}+\mu\frac{dP}{d\mu}-P
\end{align}
is no longer satisfied. 
To see this more clearly, one has to remember some thermodynamical and statistical concepts. 
For a system described by a Hamiltonian $H$, the pressure, energy density and the conserved charged number density are given by the relations
\begin{align}
P=&\frac{T}{V}\ln Z=\frac{T}{V}\ln \operatorname{tr}\left(e^{-(H-\mu N)/T}\right),\\
 \mathcal{E}=&\frac{1}{V}\frac{1}{Z}\operatorname{tr}\left(H e^{-(H-\mu N)/T}\right),\label{Eden}\\
 \rho=&\frac{1}{V}\frac{1}{Z}\operatorname{tr}\left(N e^{-(H-\mu N)/T}\right),\label{rho}
\end{align}
where $Z$ represents the partition function  while $V$ represents the volume. 
The thermodynamic relations are
\begin{align}
 \mathcal{E}=&TS+\mu\rho-P,\\
 S=&\left(\frac{\partial P}{\partial T}\right)_\mu,\label{entro}\\
 \rho=&\left(\frac{\partial P}{\partial \mu}\right)_T\label{rhodpdt},
\end{align}
where $S$ is the entropy density.
As explained in Ref. \cite{Gorenstein:1995vm}, if the Hamiltonian has an in-medium dependence throughout some parameter $w(T,\mu)$, the thermodynamic definitions, Eqs. (\ref{entro}) and (\ref{rhodpdt}), are in contradiction with the definitions given by Eqs. (\ref{Eden}) and (\ref{rho}) due to the extra terms represented by $\frac{\partial H}{\partial w}\frac{\partial w}{\partial T}$ or $\frac{\partial H}{\partial w}\frac{\partial w}{\partial \mu}$. 
In other words, the energy density and the number density can lose their meaning of representing the expected value of the energy and the number of particles respectively due to the fact that  the Hamiltonian depends on external control parameters. 
To circumvent this problem, it is necessary to define an effective Hamiltonian such as
\begin{align}
 H_{eff}=H+E(T,\mu),
\end{align}
where $E(T,\mu)$ has the following properties,
\begin{align}
 \frac{\partial E}{\partial T}=&- \frac{\partial H}{\partial w} \frac{\partial w}{\partial T},\\
 \frac{\partial E}{\partial \mu}=&- \frac{\partial H}{\partial w} \frac{\partial w}{\partial \mu},
\end{align}
so that Eqs. (\ref{entro}) and (\ref{rhodpdt}) are no longer in conflict with Eqs. (\ref{Eden}) and (\ref{rho}). 
Note that such a procedure is clearly equivalent to  a similar redefinition of the pressure\footnote{It is important to have in mind that this procedure is to be applied to the pressure evaluated with the optimized quantities, $P(\bar{\eta}(\mu),g(\Lambda),\bar {B}_2(\mu),\mu)$}. 
In our specific case ($T=0$)  the implicit parameter dependence on $\mu$ appears through $\bar\eta$, $\Lambda$ and $g(\Lambda)$, recalling that $\Lambda$ is chosen to be  $\mu$-dependent. 
Then, the thermodynamic consistent pressure has the general form \cite{Gorenstein:1995vm},
\begin{align}
 P_b(\mu)=P(\mu)+b[\eta(\mu),\Lambda(\mu),g(\Lambda)].
 \label{termo}
\end{align}
 Having settled the thermodynamic consistency issue let us start by considering the massless NNLO pQCD pressure \cite{Fraga2001,Fraga:2001xc,Vuorinen:2003fs},
 \begin{align}
\begin{split}
\frac{P^{\rm{pQCD}}}{P_{\rm{fq}}} =& 1 - 2\frac{g}{4\pi^2}  -\frac{g^2}{\left(4\pi\right)^2} \left [ 10.3754- 0.535832 N_f+N_f\ln(N_f)+ N_f \ln (\frac{g}{4\pi^2})+ \left(11-\frac{2}{3}N_f\right) \ln \left (\frac{\Lambda}{\mu_f} \right ) \right]   \; ,
 \label{pQCD}
 \end{split}
\end{align}
 where 
 \begin{align}
  P_{\rm{fq}}=N_c\frac{\mu_f^4}{12\pi^2}\,,
 \end{align}
is the free quark pressure {\it per flavor}. 
Then, according to Eq. (\ref{termo}), the thermodynamic consistent pQCD pressure can be written as
\begin{align}
 P^{\rm pQCD}_b(\mu_f)=P^{\rm pQCD}(\mu_f)+b^{\rm pQCD}(\mu_f),
\end{align}
where
\begin{align}
b^{\rm pQCD}(\Lambda(\mu_f))=-B^{\rm pQCD}-\int_{\mu_0}^{\mu_f}d\mu^\prime\frac{\partial \Lambda(\mu^\prime)}{\partial \mu^\prime}\left(\frac{\partial P^{\rm pQCD}(\mu^\prime)}{\partial \Lambda}+\frac{\partial P^{\rm pQCD}(\mu^\prime)}{\partial g}\frac{\partial g(\Lambda)}{\partial \Lambda}\right), \label{bpqcd}
\end{align}
with $\mu_0$ representing the chemical potential at which $\rho(\mu_0)=0$ while $B$ represents an integration constant which plays a role similar to that played by the bag constant in the MIT bag model \cite{Chodos:1974pn}.
Now, considering Eq. (\ref{rhodpdt}) one can write the quark number density   as\footnote{Eqs (\ref{pQCD}) and (\ref{rhopQCD}) are written differently from those appearing in Refs. \cite{Fraga2001,Fraga:2001xc,Kurkela:2009gj} in order to better fit our notation.} \cite{Kurkela:2009gj}
\begin{align}
\begin{split}
 \frac{\rho^{\rm pQCD}(\mu_f)}{\rho_{\rm{fq}}}=&1- 2\frac{g}{4\pi^2}  -\left(\frac{g}{4\pi^2}\right)^2\left [ 7.62538 -0.369165 N_f+N_f\ln(N_f)+ N_f \ln(\frac{g}{4\pi^2}) +\left(11-\frac{2}{3}N_f\right)  \ln \left (\frac{\Lambda}{\mu_f} \right ) \right] ,
 \end{split}\label{rhopQCD}
\end{align}
where the free quark number density {\it per flavor} is given by
\begin{align}
 \rho_{\rm{fq}}=N_c\frac{\mu_f^3}{3\pi^2}.
\end{align}

Similarly, for the RGOPT we have 
\begin{align}
 P_{b}^{RGOPT}(\bar\eta,g,\bar{B}_2,\mu_f)=&  P^{\rm{RGOPT}}_{1,f}(\bar\eta,g,\bar{B}_2,\mu_f)+b^{\rm RGOPT}(\mu_f),\\
 b^{\rm RGOPT}(\mu_f)=&-B^{\rm RGOPT}-\int_{\mu_0}^{\mu_f}d\mu^\prime\bigg [\frac{\partial \Lambda(\mu^\prime)}{\partial \mu^\prime}\left(\frac{\partial P^{\rm{RGOPT}}_{1,f}}{\partial \Lambda}+\frac{\partial P^{\rm{RGOPT}}_{1,f} }{\partial g}\frac{\partial g(\Lambda)}{\partial \Lambda}\right)+4\bar\eta^4gs_0\frac{\partial \bar{B}_2(\mu^\prime) }{\partial \mu^\prime}\bigg]. \label{brgopt}
\end{align}
Applying Eq(\ref{rhodpdt}) to the NLO RGOPT pressure one gets the following general expression for the quark number density 
\begin{align}
 \rho^{\rm{RGOPT}}_{1,f}=&\frac{\partial P^{\rm{RGOPT}}_{1,f}}{\partial \mu_f}\notag\\
 =&N_c\left(\frac{p_F^3}{3\pi^2}+\bar\eta^2\frac{\gamma_0}{2b_0}\frac{p_F}{\pi^2}\right)-\frac{ g d_A}{\left(2\pi\right)^4}p_F\left(2\bar\eta^2+\mu_f p_F-3\bar\eta^2 L_\mu\right).
\end{align}
Let us stress that, alternatively, thermodynamic consistency can be achieved by integrating the density, $\rho$, to obtain a thermodynamic consistent pressure as done for instance in Ref. \cite{Kurkela:2009gj}. In principle, this approach must be equivalent to the one presented here. Nevertheless, in our specific case, all the RGOPT prescription is derived from the pressure, and in this sense one needs this quantity to be the starting point. Let us also remark  that the important problem related to thermodynamic consistency has been considered in many other studies (see Refs. \cite{Gorenstein:1995vm,Wang:2000dc,Yin:2008me,Lenzi:2010mz,Xia:2014zaa,Xu:2014zea}). Finally, regarding the potential scale dependence one should note that, contrary to its  RGOPT counterpart, the pQCD pressure, Eq. (\ref{pQCD}), does not carry any RG information (through RG coefficients such as $\gamma_0$, $b_0$, etc). Therefore, it 
should come as no surprise that, in general, pQCD results  present a stronger scale dependence than the RGOPT ones.
\section{Results}\label{results}
Let us now obtain some numerical results for the EoS and mass-radius relations for NSQS starting with the equation of state with charge neutrality and  in chemical equilibrium.
\subsection{Equation of state in chemical equilibrium}
The conditions for chemical equilibrium and charge neutrality conditions are respectively given by
\begin{align}
 \mu_u=\mu_d-\mu_e\equiv \mu, \label{betaeq}\\
 \frac{2}{3}\rho_u-\frac{1}{3}\rho_d-\rho_e=0,\label{charge_neu} \,
\end{align}
where $\rho_e=\mu_e^3/(3\pi^2)$ is the number density for free massless electrons. Eqs. (\ref{betaeq}) and (\ref{charge_neu}) allow us to fix $\mu_d(\mu)$ and $\mu_e(\mu)$.
In the case of  asymmetric quark matter it can be  convenient to express the thermodynamic quantities in terms of the baryon chemical potential, $\mu_B$,   defined as
\begin{align}
    \mu_B=\mu_u+2\mu_d.
\end{align}
\begin{figure}
 \centering
 \includegraphics[width=.6\textwidth]{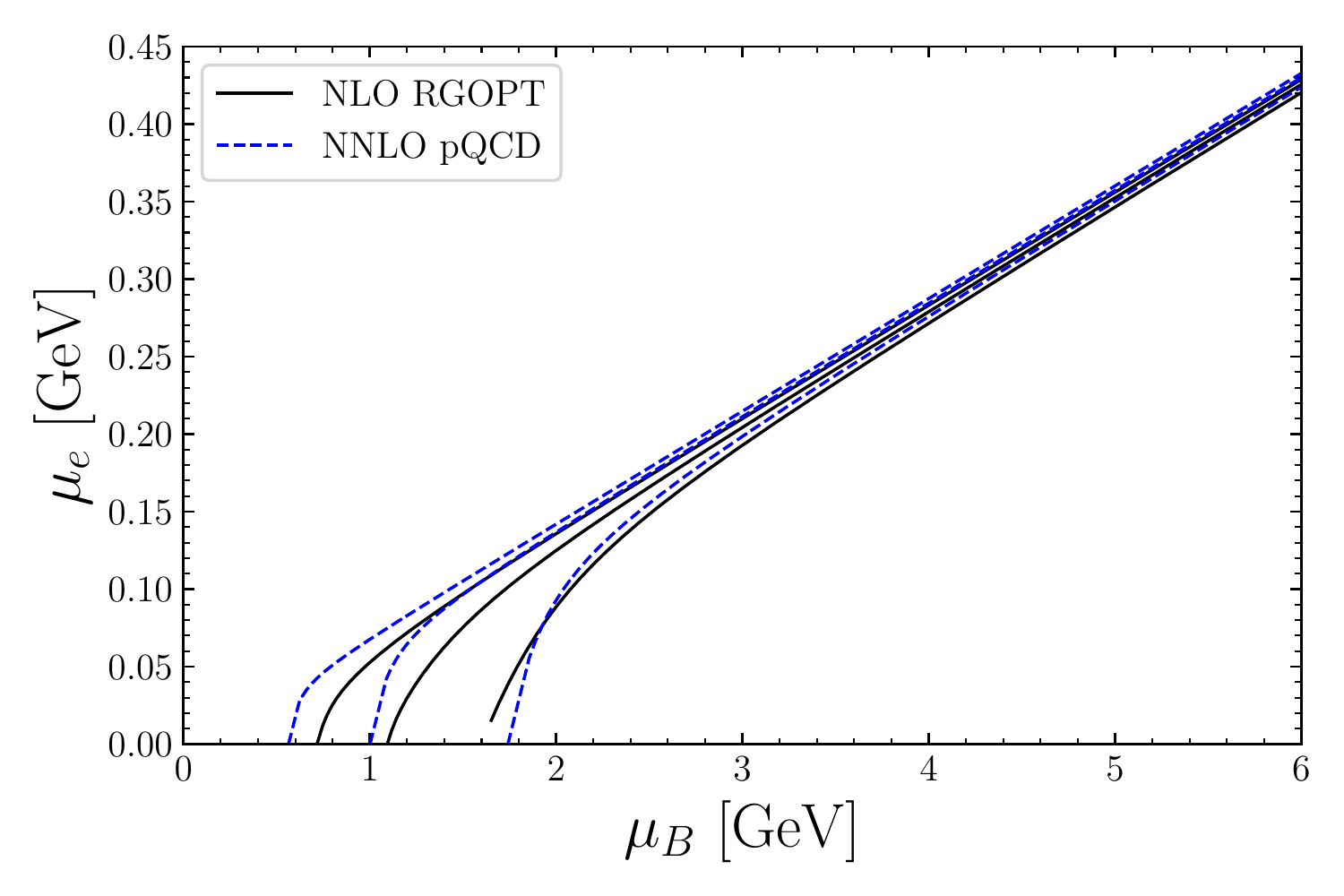} 
\caption{Electron chemical potential, as a function of $\mu_B$, obtained from the charge neutrality condition, Eq. (\ref{charge_neu}). The NLO RGOPT  is represented by the continuous line  and the NNLO pQCD by the dashed line. The boundaries of each band are obtained by setting the renormalization scale coefficient to $X=1$ (bottom boundary) and $X=4$ (top boundary), while $X=2$ corresponds to the central line.}
\label{muevsmu_B0}
\end{figure}
In Fig. \ref{muevsmu_B0}, the electron chemical potential is plotted as a function of the baryon chemical potential. As $\mu_B$ increases, $\mu_e$ also increases as expected from the charge neutrality and chemical equilibrium constraints. 
The figure shows that both approximations  predict a similar electron production at $\mu_B\gtrsim 3$ GeV. 
Note also that, within this asymmetric quark matter, the electron contribution to the pressure and to the density is negligible for the most relevant $\mu_B$ range.
As already discussed in Ref. \cite{Glendenning:2000}, charge neutrality is achieved more favorably by charge carrying baryons, since baryon number is conserved while lepton number is not conserved because neutrinos leave the star due to their very large mean-free path. 
In NSQM, charge neutrality is easily attained by considering matter with 1/3 $u$-quarks and 2/3 $d$-quarks. 
For instance, in Ref \cite{CamaraPereira:2016chj}, where several NJL like models were discussed, including versions with a vector interaction with different  magnitudes, a non-zero electron contribution could only be achieved with a quite large vector term, but even then the resulting electron fraction is $ \lesssim 0.005$.
It can be easily checked that in our case, the electron pressure is at least two orders of magnitude lower than the total pressure for the most relevant $\mu_B$ range. 
The electron pressure is comparable to the quark pressure only near the $\mu_e(\tilde{\mu}_B)$ value, where $P(\tilde{\mu}_B) = 0$, while the electron density is quite small when compared with the quark number density for all $\mu_B$ values. 
Then, the total pressure and densities are 
\begin{align}
\begin{split}
 P=P_q+P_e,\\
 \rho=\rho_q+\rho_e,
\end{split}
\end{align}
where $P_e=\mu_e^4/(12\pi^2)$ is the pressure of a gas of free (massless) electrons. 
In all results presented here, we shall consider $B=0$ and $\Lambda=X(\mu_u+\mu_d)/2$, where the renormalization scale coefficient, $X$, will initially take the values 1, 2 and 4,  with 2 representing the so-called central value.
\begin{figure}
\begin{subfigure}
  \centering
 \includegraphics[width=.45\textwidth]{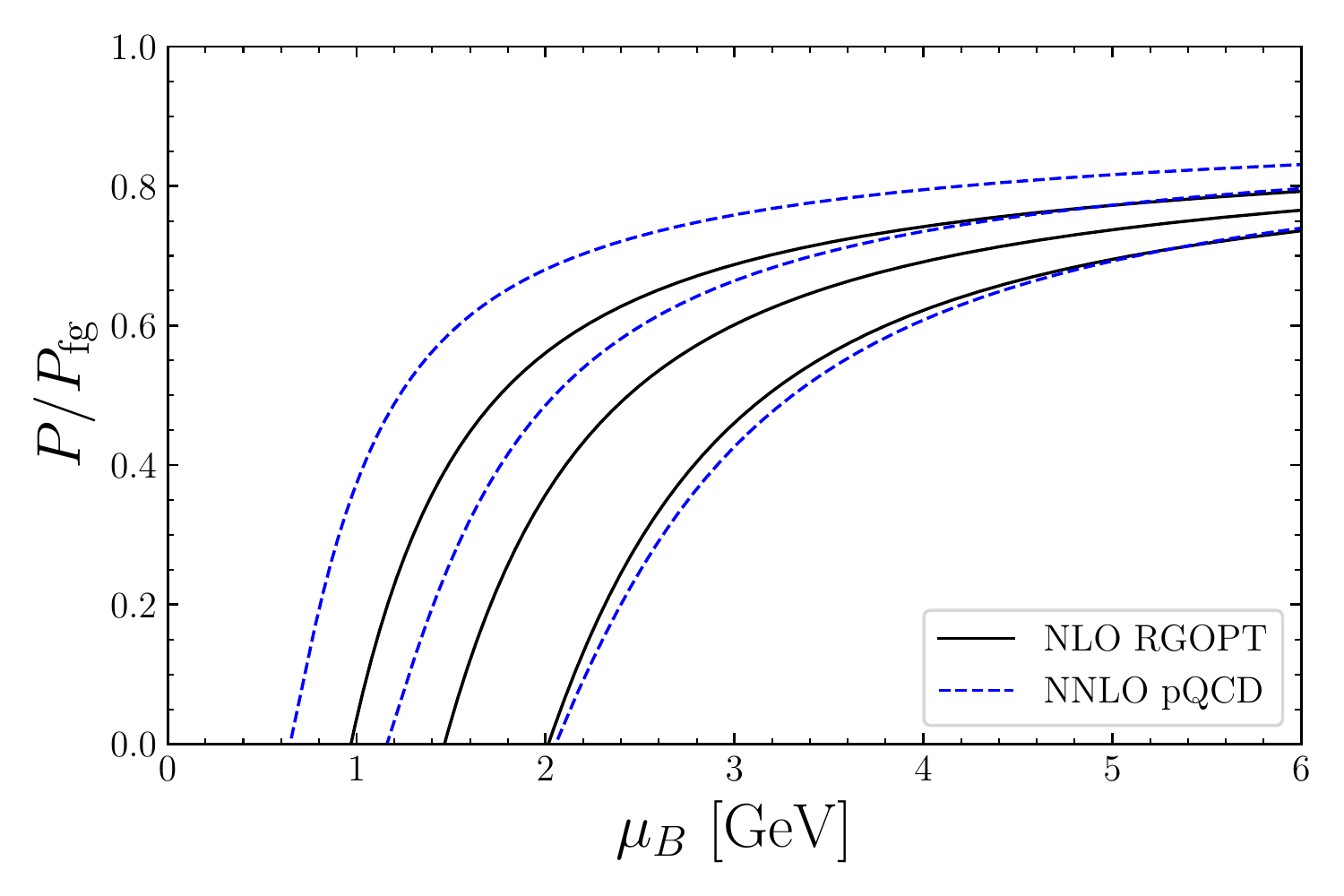}
\end{subfigure} 
\begin{subfigure}
  \centering
 \includegraphics[width=.45\textwidth]{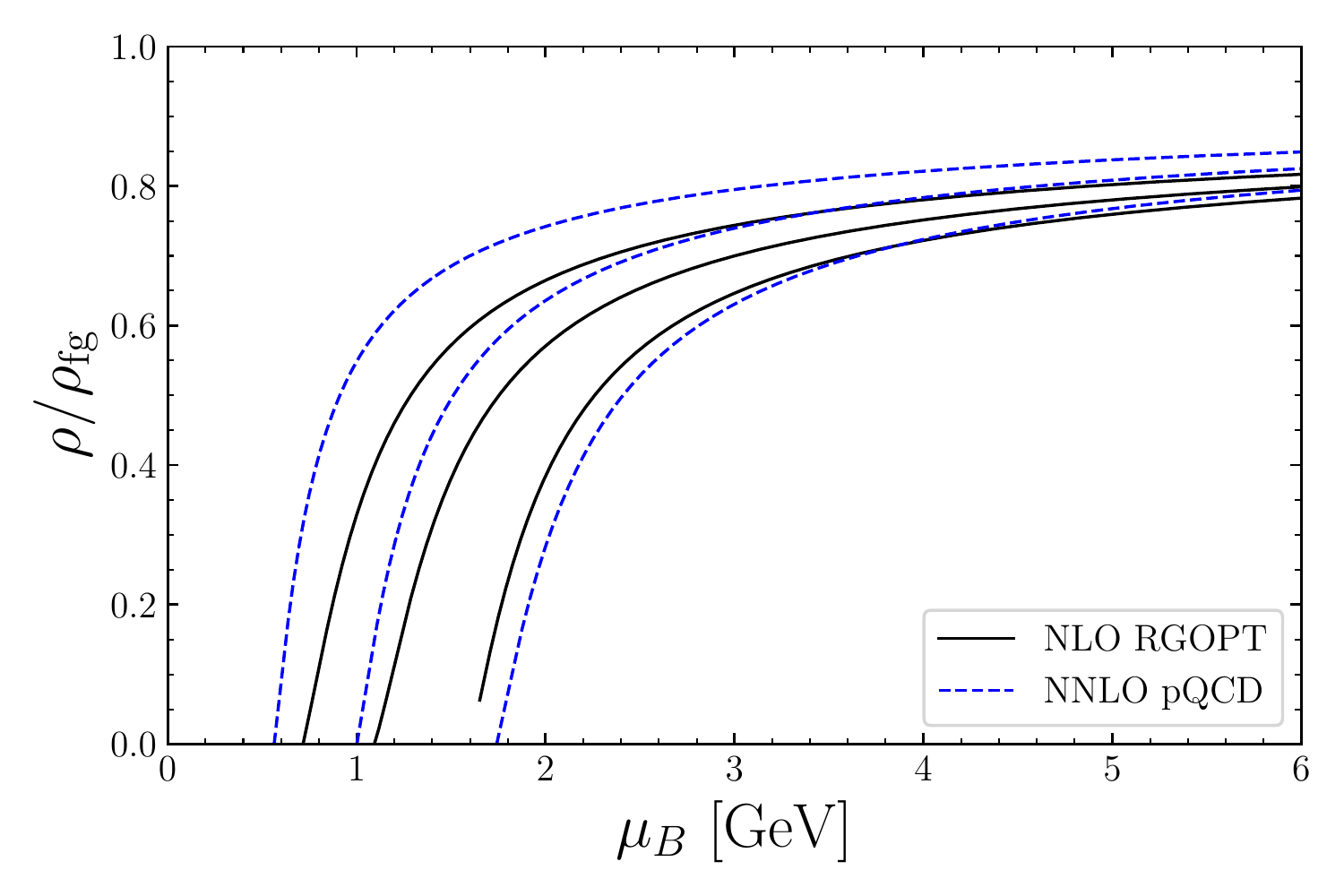}
\end{subfigure} 
\caption{Thermodynamic consistent pressure (left) and quark number density (right), as functions of baryon chemical potential, furnished by the RGOPT at NLO (continuous line) and by pQCD at  NNLO  (dashed line). The boundaries of each band are obtained by setting  the renormalization scale coefficient to $X=1$ (bottom boundary) and $X=4$ (top boundary), while $X=2$ corresponds to the central line.}
\label{P_rho_ther_consis_B0}
\end{figure}
In Fig. \ref{P_rho_ther_consis_B0} we present the total pressure and the total quark number density as functions of the baryon chemical potential.
In general, the RGOPT produces a softer EoS than the one  furnished by pQCD allowing us to  anticipate that it will produce lower maximum NSQS masses. 
For $X=1$, the value of $\mu_B$ at which $P=0$ is $\tilde{\mu}_B=2.01(2.05)$ GeV,  at $\alpha_s\sim 0.70(0.68)$, for the NLO RGOPT(NNLO pQCD) while the baryon density is $\rho_B=\rho_q/3\sim 6.2(5.3)\rho_0$ where $\rho_0= 0.16 \ \text{fm}^{-3}$ is the saturation density of nuclear matter.
For $X=2$, $\tilde{\mu}_B=1.46(1.16)$ GeV,  at $\alpha_s\sim 0.45(0.58)$, while the baryon density is $\rho_B \sim 2.2(0.7)\rho_0$. 
Now, for $X=4$, $\tilde{\mu}_B=0.970(0.648)$ GeV at $\alpha_s\sim 0.36(0.51)$, while the baryon density is $\rho_B\sim0.5(0.1)\rho_0$ implying that in order to describe realistic NSQM one needs to restrict the values of $X$.

\begin{figure}
 \centering
 \includegraphics[width=.6\textwidth]{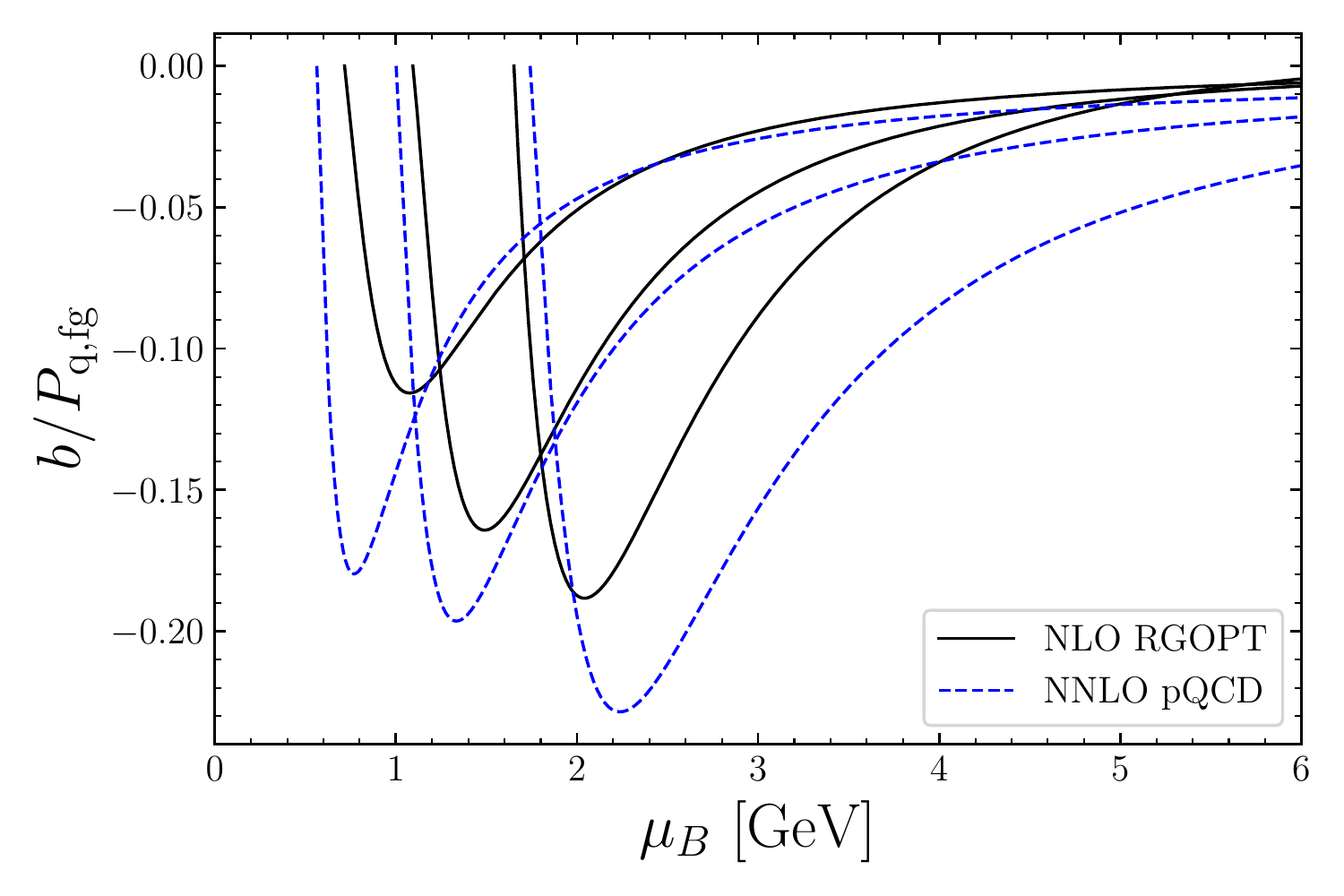} 
\caption{$b$ as a function of the baryon chemical potential  for  the RGOPT at NLO (continuous line) and  pQCD at NNLO (dashed line). The boundaries of each band are obtained by setting  the renormalization scale coefficient to $X=1$ (bottom boundary) and $X=4$ (top boundary),while $X=2$ corresponds to the central line. }
\label{b_B0}
\end{figure}
Fig. \ref{b_B0}, which  shows the value of $b/P_fg$ as a function of the baryon chemical potential, indicates that $b$ can be interpreted as a measure  of how far $P_{1,f}^{\rm{RGOPT}}$ and $P^{\rm{pQCD}}$ are from being thermodynamic consistent. 
The results show that the impact of $b$ in the pressure is more important at lower $\mu_B$ values, affecting considerably the properties of the resulting QS mass-radius relation.
By noting that the term in parenthesis in Eqs (\ref{bpqcd}) and (\ref{brgopt}) is just the reduced RG equation times $1/\Lambda$ one may conclude that the thermodynamic inconsistency can also be treated by considering the RG properties displayed by the relevant perturbative expressions. 
In general, the absolute value of $b$ is lower for the RGOPT and, as $\mu_B$ increases, approaches zero faster than the pQCD result.
\subsection{Non-strange quark stars}
\begin{figure}
\begin{subfigure}
    \centering
     \includegraphics[width=.45\textwidth]{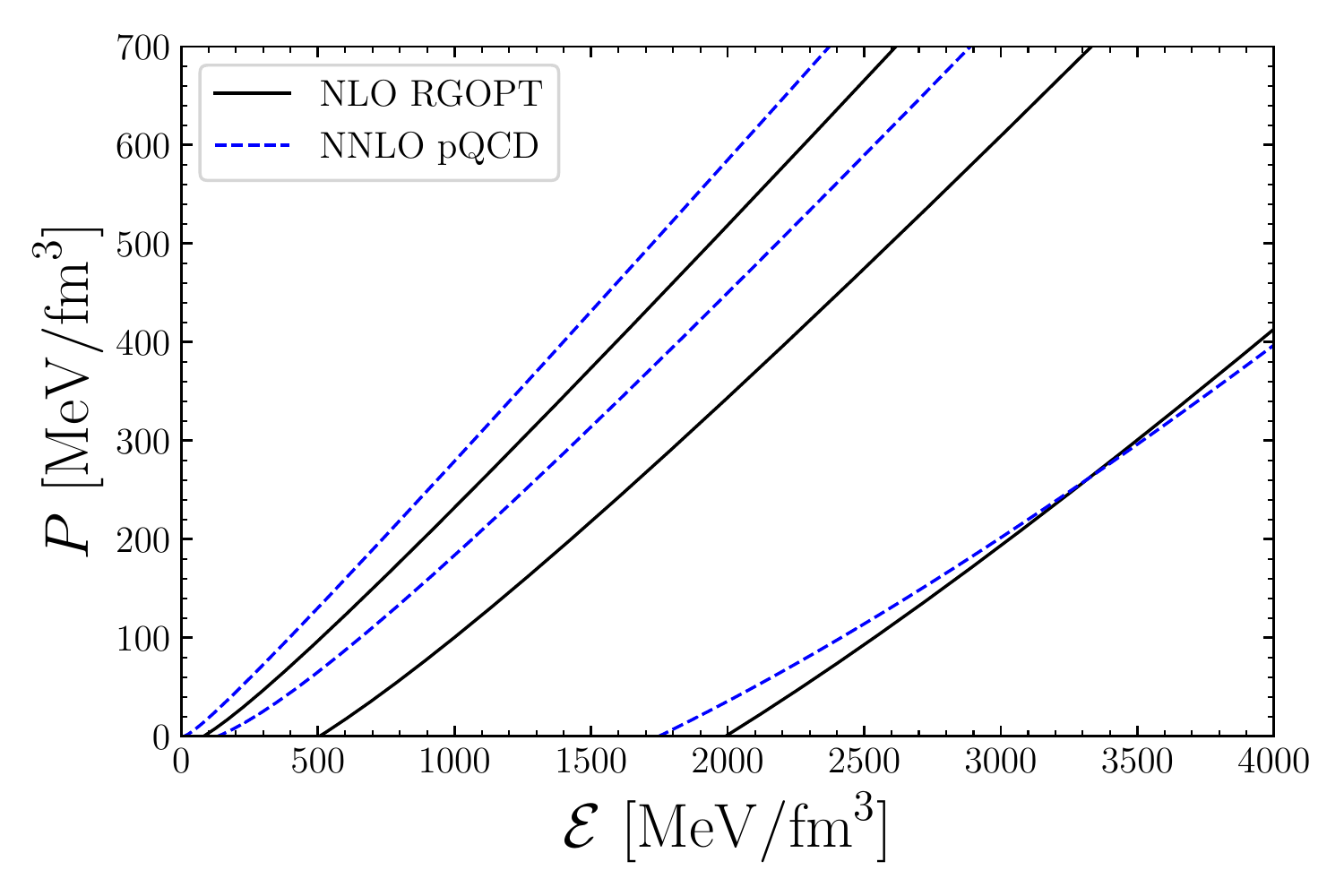} 
\end{subfigure}
\begin{subfigure}
    \centering
 \includegraphics[width=.45\textwidth]{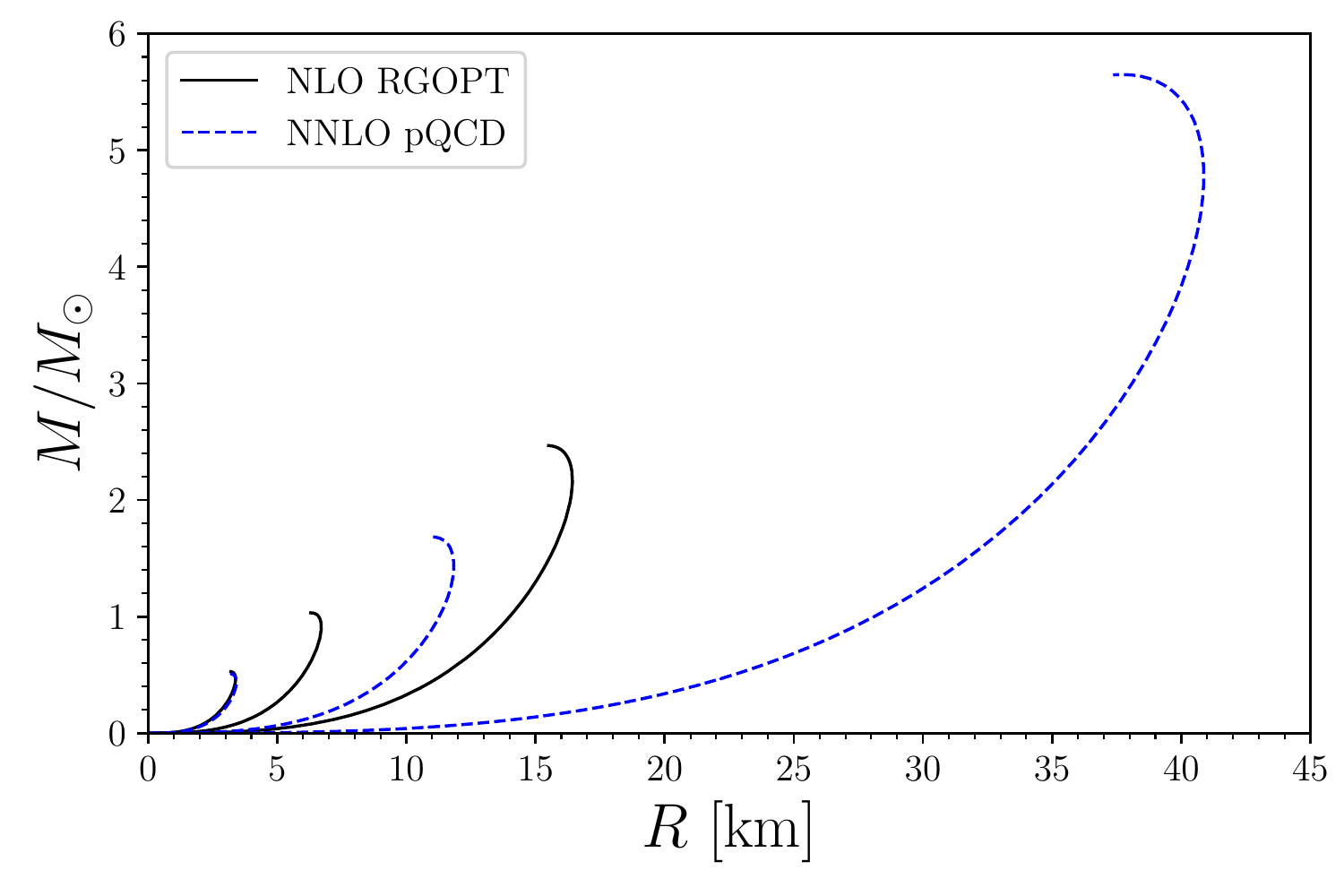} 
\end{subfigure}
\caption{Left panel: EoS furnished by the  RGOPT at NLO (continuous line) and  by pQCD at NNLO (dashed line). The boundaries of each band are obtained by setting  the renormalization scale coefficient to $X=1$ (bottom boundary) and $X=4$ (top boundary), while $X=2$ corresponds to the central line. Right panel: Mass-radius relation given by the EoS plotted in the left panel. The curves representing the higher maximum masses correspond to $X=4$ and the ones representing lower maximum masses correspond to $X=1$ for both approximations.}
\label{MvsR_B0}
\end{figure}

In general relativity, the mass-radius relation for a non-rotating hydrostatic compact star is determined by solving the Tolman--Oppenheimer--Volkoff equations \cite{Tolman:1939jz,Oppenheimer:1939ne}
\begin{align}
\begin{split}
 dP(r)=&-\frac{G\left(P(r)+\mathcal{E}(r)\right)\left(M(r)+4\pi r^3P(r)\right)}{r\left(r-2GM(r)\right)}dr,\\
 dM(r)=&4\pi r^2\mathcal{E}(r)dr,
\end{split}
\label{TOV}
\end{align}
where $G=(1.22\times 10^{19})^{-2} \ \text{GeV}^{-2}$ is Newton's gravitational constant and $r$ is the radial coordinate of the star.
The EoS  enters through $\mathcal{E}(P)$ (see Fig. \ref{MvsR_B0}) while Eqs. (\ref{TOV}) are solved by choosing a value for the pressure at the center of the star, $P(r=0)$ followed by an outward  integration to the surface where the pressure is zero.
\begin{table}
\caption{QS properties predicted by the NLO RGOPT and NNLO pQCD for three values of $X$. }
\begin{center}
\begin{tabular}{|c  |c  c  c   c c|} 
       
       \hline
    &  $X$ & $M_{\rm max}/M_\odot$ & $R_{\rm max}$ [km] & $\rho_B^{c,\rm max}$ & $\mu_B^c$ [GeV]  \\
	\hline
	\multirow{2}{4.5em}{NLO RGOPT $\ \ $}&$1$& $0.53 $ & $ 3.20$  & $32.8\rho_0$ & $2.987$  \\
    & $2$ & $1.03 $ &  $6.30$ &  $11.9\rho_0$ & $2.179 $ \\ 
  
     &$4 $& $2.47 $ & $ 15.5$  & $2.90\rho_0$ & $1.408$  \\
     \hline
       \multirow{2}{4.5em}{NNLO pQCD $\ \ $}&$1$ & $0.51 $ & $3.19$ & $33.5 \rho_0 $ & $3.028$\\
       & $2$ & $1.68$ & $11.0$ & $ 4.94 \rho_0$ & $1.670$  \\
          
     &$4$ & $5.65 $ & $30.7$ & $0.77 \rho_0 $ & $0.926$\\
	\hline
\end{tabular}\label{tab:tabla4.1a}
\end{center}
\end{table}
The EoSs plotted in the left panel of Fig. \ref{MvsR_B0} produce the mass-radius relations displayed in Fig. \ref{MvsR_B0}
clearly indicating how the RGOPT  promotes a substantial improvement regarding  scale dependence, when compared to pQCD.
 In Table \ref{tab:tabla4.1a}, we give the values of some relevant quantities describing the most massive quark star as predicted by both approximations for three values of $X$ ($X=1,2,4$) so that one can compare the two approaches at a given renormalization scale. Notice that, for the same scale,  the RGOPT predicts lower maximum masses and lower radii when compared to pQCD, with the exeption of the lower scale value ($X=1$) when the masses and radii are almost identical, $M_{\rm{max}}\approx 0.51M_\odot$. 
 More precisely, the RGOPT produces   quark stars whose maximum masses and radii are  $M_{\rm max}(X=2)\approx 1.03 M_\odot$ and  $R_{\rm{max}}\approx 6.3$ km,  $M_{\rm max}(X=4)\approx 2.47 M_\odot$ and $R_{\rm{max}}\approx 16$ km while pQCD produces the following masses and radii values, $M_{\rm max}(X=2)\approx 1.68 M_\odot$ and $R_{\rm{max}}\approx 11$ km,  $M_{\rm max}(X=4)\approx 5.65 M_\odot$ and $R_{\rm{max}}\approx 38$ km.

Then, varying the value of $X$ from 1 to 4, pQCD produces a variation in the maximum mass of $5.14M_\odot$ while the RGOPT induces a variation of $1.97 M_\odot$ emphasizing, once again, the RGOPT smaller sensitivity to  scale changes. 
Furthermore, it is interesting to notice that  the value $\sim 2.5\, M_\odot$ for the maximum mass, at   $X=4$, is in accordance with the mass of a compact stellar object  recently remarked by the LVC. 
Namely, the detection of  event GW190814 \cite{LIGOScientific:2020zkf} with a mass $2.50-2.67M_\odot$, which is still not discarded to be a NSQS \cite{Cao:2020zxi} \footnote{At the moment, it is not clear if this object is a neutron/quark star or a black hole \cite{LIGOScientific:2020zkf}, since its mass is  near black hole candidates. Therefore, it could be a light black hole or a very massive NS or QS}.
It should also be obvious that, by choosing an appropriate value for $X$, pQCD is also able to predict values $M_{\rm max}\leq 2.6$. 
Nonetheless, by taking into account the current observational data and the theoretical uncertainty set by $X=1-4$, one becomes convinced that the RGOPT predictions lie within a more  realistic  range than those predicted by pQCD (see Fig. \ref {MvsR_B0}, right panel). 
\subsection{Constraining the non-strange quark matter equation of state}
In order to obtain realistic mass-radius relations one must constrain the values of $X$ such that the RGOPT and pQCD predict maximum star masses compatible with the actual observational data. 
Then, we can restrict the values of $X$ for both approximations, such that the maximum star masses given by the TOV equations lie between the usual bound ($M_{\rm{max}}\geq 2M_\odot$) and the mass of the low compact object associated with GW190814, so that $2M_\odot\leq M_{\rm{max}}(X)\leq 2.6 M_\odot$. Notice also that in Ref. \cite{Annala:2021gom} a maximum NS mass of 2.53$M_\odot$ was predicted, considering that a  black-hole was formed associated with the GW170817 after the decay of the supermassive  NS that resulted from the merging, i.e., it has a mass above the maximum mass predicted by integrating the TOV equations, generally designated as $M_{TOV}$. Therefore, we consider it reasonable to take the    maximum mass as $M_{\rm{max}}\sim 2.5-2.6 M_\odot$. This is also the maximum mass predicted by several Bayesian studies which consider a relativistic mean field description in conjunction with a reasonable set of constraints \cite{Malik:2022zol,Malik:2023mnx}. The minimum mass  is chosen by taking into account the recent observational data of two solar mass pulsars \cite{Demorest:2010bx,Antoniadis:2013pzd,Romani:2021xmb}. In particular, the pulsar PSR J1810+1744 whose mass is 2.13$\pm 0.04 M_\odot$ lies above the limit of two solar masses within 3$\sigma$ \cite{Romani:2021xmb}.

Imposing these  maximum mass constraints, we found that for the RGOPT calculation the renormalization scale coefficient is $3.41\lesssim X\lesssim 4.17$, while for pQCD, it is $2.21\lesssim X\lesssim 2.57$, see Table \ref{tab:tabla4.1}. These values, apart from the evident improvement in the scale dependence of the RGOPT, also indicate that, within the relevant $\mu_B$ range, the RGOPT operates at lower $\alpha_s$ values than pQCD. In other words, since the RGOPT considers  higher $X$ optimal values (lower values of $\alpha_s$), its results are in the range where  perturbative calculations become more reliable. In Table \ref{tab:tabla4.1}, we also present other properties as the radius of the star with maximum mass, the corresponding central baryonic density and chemical potential, as well as the masses of stars with  masses  1.4$M_\odot$ and 0.77$M_\odot$ (see the discussion below).

\begin{figure}
\begin{subfigure}
     \centering
     \includegraphics[width=.45\textwidth]{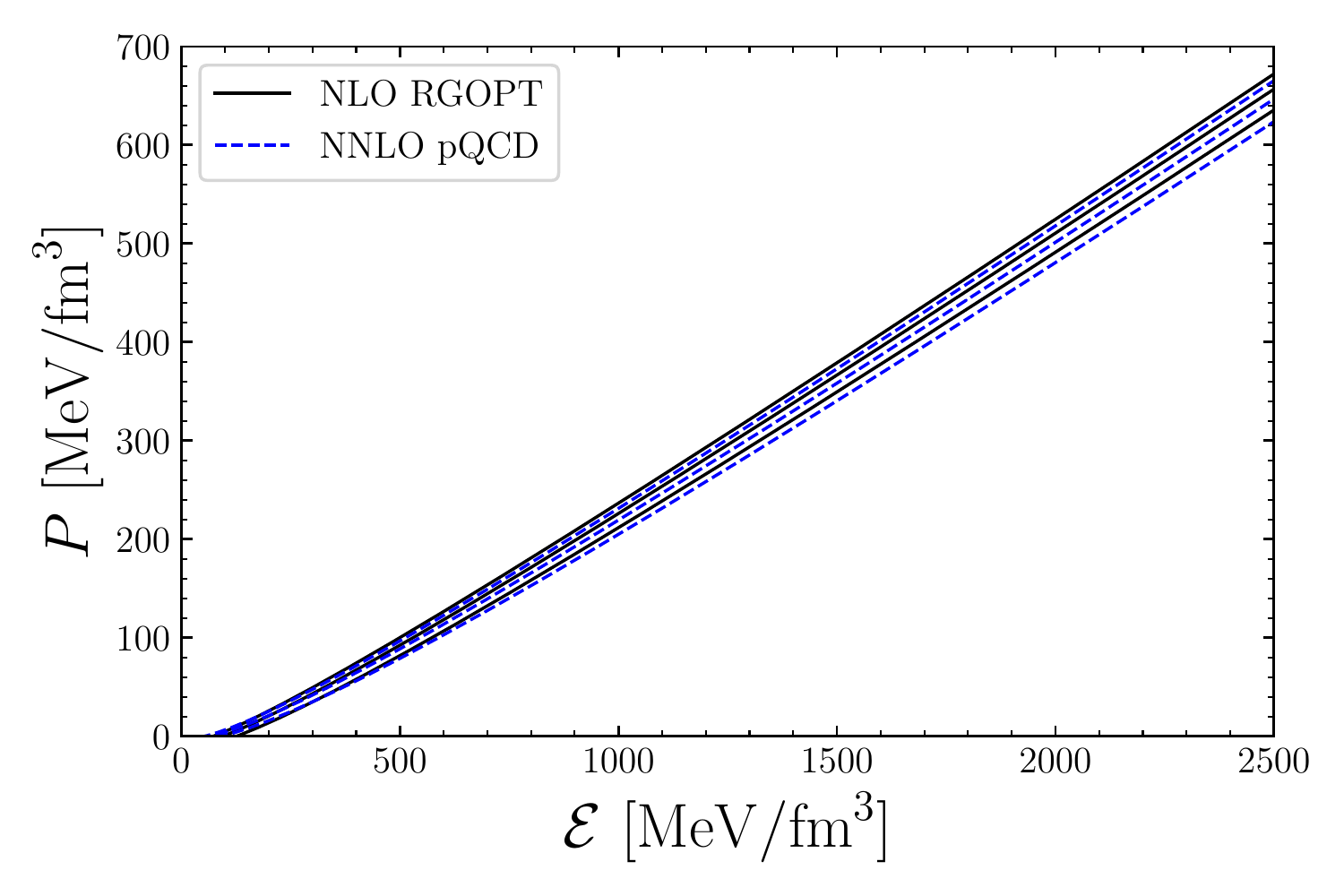} 
\end{subfigure}
 \centering
 \includegraphics[width=.45\textwidth]{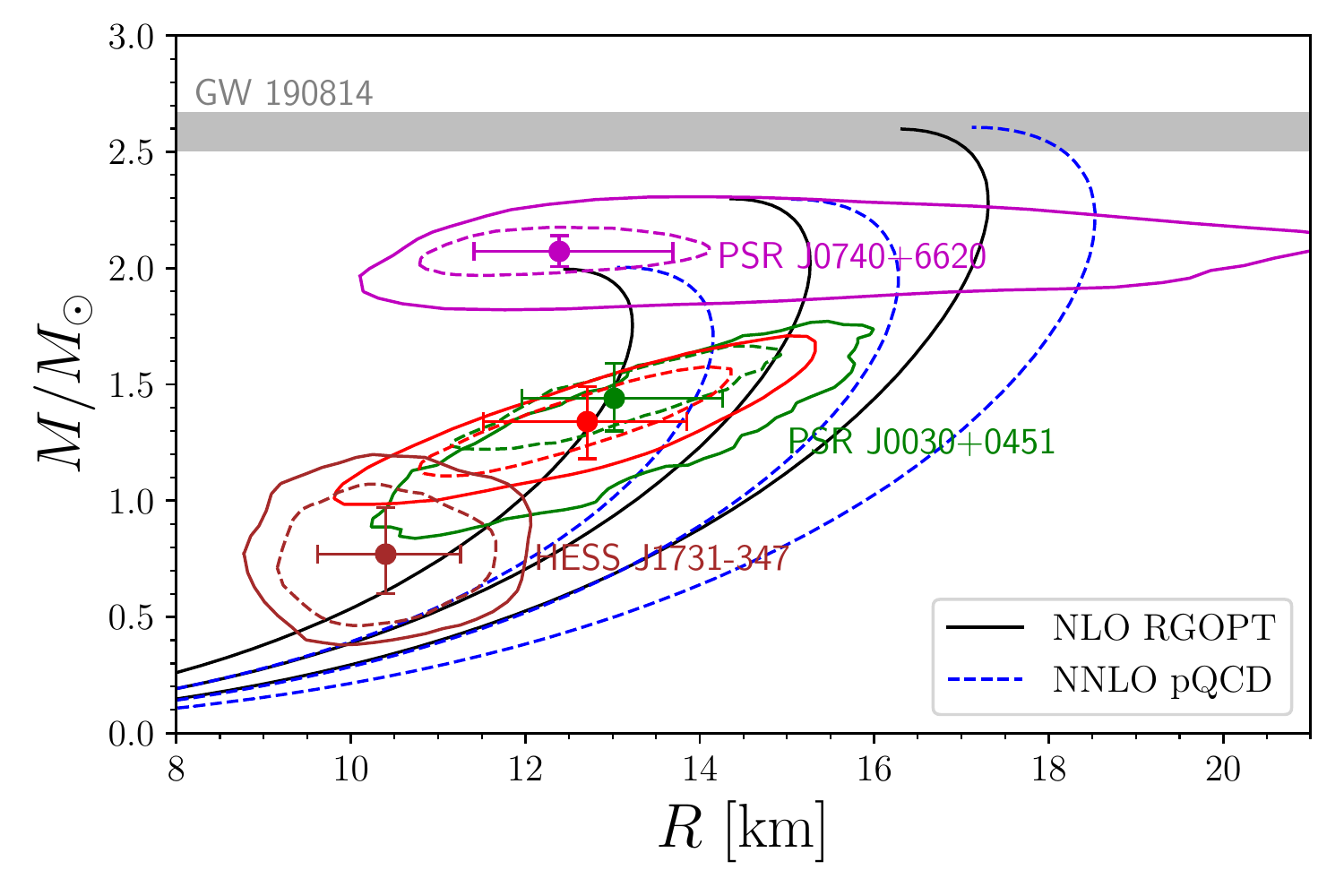} 
\caption{Left panel: Functional dependence of the pressure and the energy density for the NLO RGOPT (full line) and NNLO pQCD (dashed line). The central lines correspond to $X=3.81$ for the NLO RGOPT and $X=2.39$ for the NNLO pQCD. The external lines correspond to $X=3.41$ (bottom) and $X=4.17$ (top) for the RGOPT, while for pQCD, $X=2.21$ (bottom) and $X=2.57$ (top). Right panel: Mass-radius relation given by the EoS plotted in the left panel.  Also included are  data from NICER for the pulsars PSR J0030+0451 \cite{Riley_2019,Miller:2019cac} and PSR J0740+6620 \cite{Riley:2021pdl,Miller:2021qha},  the low mass compact star HESS J1731-347 \cite{2022NatAs.tmp..224D}. In particular, the ellipses represent the 68\% (dashed) and  95\% (full) confidence interval of the 2-D  distribution in the mass-radii domain  while the error bars give the 1-D marginalized posterior distribution for the same data. A band identifying the mass of the low mass compact object associated with GW190814 \cite{LIGOScientific:2020zkf} has also been included. }
\label{MvsR_B0_Aopt_band}
\end{figure}
\begin{table}
\caption{QS properties, maximum mass $M_{\rm max}$ and corresponding radius $R_{\rm max}$,  radius  of the 1.4$M_\odot$ and 0.77$M_\odot$ stars, $R_{1.4}$ and $R_{0.77}$, central baryon density $\rho_B^{c,\rm max}$ and central baryon chemical potential $\mu_B^c$ of maximum mass configurations predicted by the NLO RGOPT and NNLO pQCD for maximum mass stars 2.0, 2.3 and 2.6 $M_\odot$.}
\begin{center}
\begin{tabular}{|c  |c  c  c  c c c c|} 
       \hline
    &  $X$ & $M_{\rm max}/M_\odot$ &  $R_{\rm max}$ [km] & $R_{1.4}$ [km] & $R_{0.77}$ [km] & $\rho_B^{c,\rm max}$ & $\mu_B^c$ [GeV]  \\
	\hline
	\multirow{2}{4.5em}{NLO RGOPT $\ \ $} & $3.41$ & $2.00 $ &  $12.4$ & $12.9$ & $11.1$& $4.13\rho_0$ & $1.569 $ \\ 
  &$3.81$& $2.30 $ & $ 14.4$ &$14.4$ & $12.4$& $3.33\rho_0$ & $1.467$  \\
     &$4.17 $& $2.60 $ &$16.3$ & $ 15.8$ &$13.4$ & $2.69\rho_0$ & $1.376$  \\
     \hline
       \multirow{2}{4.5em}{NNLO pQCD } & $2.21$ & $2.00$ & $13.1$ & $13.9$ & $12.2$ &$ 3.95 \rho_0$ & $1.552$  \\
          &$2.39$ & $2.30 $ & $15.0$& $15.6$ & $13.4$ & $3.19 \rho_0 $ & $1.451$\\
     &$2.57$ & $2.61 $ & $17.1$ & $17.3$ & $14.8$ & $2.58 \rho_0 $ & $1.359$\\
	\hline
\end{tabular}\label{tab:tabla4.1}
\end{center}
\end{table}

The mass-radius relation curves obtained with the boundary values $M_{\rm{max}}(X)=2M_\odot$ and $M_{\rm{max}}(X)=2.6M_\odot$, and also with an intermediate value $M_{\rm{max}}(X)=2.3M_\odot$ are plotted in Fig. \ref{MvsR_B0_Aopt_band} while their main properties  are summarized in Table \ref{tab:tabla4.1}. The intermediate mass is justified because a maximum mass in the range 2.16-2.32$M_\odot$ has been predicted considering that GW170817 is associated to the formation of a hypermassive star \cite{Margalit:2017dij,Rezzolla:2017aly,Ruiz:2017due,Annala:2021gom}. Moreover, a Bayesian study \cite{Malik:2022zol} which includes the contribution of hyperons  has predicted a maximum mass of the order of 2.2$M_\odot$.

In Fig.  \ref{MvsR_B0_Aopt_band}, besides the mass-radius curves, we have also included the predicted masses and radii of the pulsars PSR J0030+0451 \cite{Riley_2019,Miller:2019cac}, PSR J0740+6620  \cite{Cromartie:2019kug,Fonseca:2021wxt,Riley:2021pdl,Miller:2021qha} and the compact object HESS J1731-347 \cite{2022NatAs.tmp..224D}, as well as the band identifying the mass of the low mass compact object associated with GW190814 \cite{LIGOScientific:2020zkf}. 

The comparison allows us to conclude that the RGOPT  results obtained with both {$M_{\rm{max}}(X)= 2M_\odot$} and {$2.3M_\odot$} are in agreement with the NICER constraints for both PSR J0030+0451 and  PSR J0740+6620 and even with  the radius predicted for the compact object  HESS J1731-347. Predictions obtained with the larger scale, however, fail to describe these objects at the 95\% confidence interval. On the other hand,  pQCD is only compatible with PSR J0030+0451 and HESS J1731-347 considering the scale $X=2.21$ corresponding to the maximum  mass, 2$M_\odot$.
In the case of $M_{\rm{max}}(X)=2.6M_\odot$, both approximations fail to describe masses and radii in the 95\% confidence interval of PSR J0030+0451 and HESS J1731-347, being only compatible with PSR J0740+6620 and GW190814.
If the low mass compact object associated with GW19081 turns out to be a NS/QS, it will be difficult to describe the  observational data of all pulsars considered,  in the case of the two approximations studied here.
The central baryonic densities of the maximum mass configurations given in Table \ref{tab:tabla4.1} for both RGOPT and pQCD are compatible with the densities one may expect to find inside NSs.

Since the renormalization scale remains a ``free'' unfixed parameter inside in-medium QCD, both approximations can predict masses between the desired range $2M_\odot\leq M_{\rm{max}}(X)\leq 2.6 M_\odot$ when the vale of $X$ is appropriately chosen. Nevertheless, the RGOPT is able to produce radii values which are in better agreement with the observational data as   Fig. \ref{MvsR_B0_Aopt_band} clearly shows. 
 
Let us also remark that the EoS can be further softened by using a finite value for $B$ so that lower maximum masses can be obtained.
In principle, to fix $B$, one must use the stability criteria of the energy per baryon $\varepsilon=E/A$, which in the case of  NSQM must be lower than 930 MeV and, at the same time, be lower than the energy per baryon of SQM \cite{Holdom:2017gdc}. 
The inclusion of a new degree of freedom to describe strangeness will also soften the EOS by reducing the Fermi pressure. Therefore, smaller maximum masses values may be expected if the same EOS parameters are used. Using observations to constrain the QCD scale will probably result in different $X_{\rm{min}}$ and $X_{\rm{max}}$, shifting these to larger values.  
This effect of the strange quark mass was already reported in pQCD \cite{Fraga:2004gz}.
The conclusions drawn in Refs. \cite{CamaraPereira:2016chj,Yuan:2022dxb}, where hybrid stars and quark stars  were studied within the SU(2) and SU(3) NJL model, also support these arguments. The   $ud$ quark  matter allows for a stiffer equation of state, and, therefore, for a higher maximum mass.

\section{Conclusions}\label{conclusion}
The RGOPT resummation method was employed in the evaluation of the QCD EoS, at NLO, in the case of NSQM cold ($T=0$) quark matter. The resulting EoS was then used  to predict the mass-radius relations for NSs and the results were compared with those furnished by standard pQCD evaluations at NNLO. 
Following the standard procedure \cite{Fraga:2004gz,Kurkela:2009gj,Kneur:2019tao}, the renormalization scale was taken to depend on the chemical potential, as $\Lambda = X (\mu_u+\mu_d)/2$, requiring a modification of the RGOPT and pQCD pressures, in order to ensure thermodynamic consistency.   
After performing the necessary adjustments we  found that thermodynamic consistency is more relevant at lower  densities,  which represents the crucial region for the description of  maximum QS masses.
Aiming to get a general idea about the stiffness and the  scale dependence  produced by both  EoS  we have started our numerical investigations by varying the scale   from the  {\it central} value, $X=2$ by a factor of two (anticipating that smaller scales give less realistic results \cite{Kneur:2019tao}). The results show that when the {\it same} scale  is considered the RGOPT produces in general a softer EoS, and therefore  smaller maximum quark star masses, than pQCD. 
As expected from previous applications \cite{Kneur:2019tao, Kneur:2021dfo,Kneur:2021feo}, we have explicitly shown that the RGOPT considerably decreases the  theoretical uncertainty given by  the scale dependence. 

This property is quite visible in the mass-radius relation, where the RGOPT predicts masses from $0.53\,M_\odot$ ($X=1$) to 
$2.47\,M_\odot$ ($X=4$) while pQCD  predicts values from $0.51\,M_\odot$ ($X=1$) to 
$5.65\,M_\odot$ ($X=4$). One should remark that at $X=4$ the RGOPT prediction is surprisingly close to the recent astronomical observations of a possible massive neutron/quark star of  $2.6\, M_\odot$ as the low mass compact object associated with the GW190814 \cite{LIGOScientific:2020zkf}. 
In contrast, pQCD behaves quite poorly at this energy scale, producing extremely high maximum quark masses, quite far from the maximum NS masses reported to  date, and with a central density well below  the saturation value.

Having identified the relevant energy scale range for the description of currently known stellar objects we have attempted to perform more realistic predictions by  constraining $X$ to describe maximum mass stars in the range $2-2.6M_\odot$. 
Adopting  this strategy we have calculated $X=3.41-4.17$ for the RGOPT and $X=2.21-2.57$ for pQCD. 
For both $M_{\rm{max}}(X)=2M_\odot$ and $2.3M_\odot$, the  radii predicted by RGOPT  ud QS are   compatible with the radius of the compact objects PSR J0030+0451, PSR J0740+6620 and HESS J1731-347, while pQCD describes these object only with $M_{\rm{max}}(X)=2M_\odot$.
The RGOPT clearly generates mass-radius relations which are less scale dependent than those furnished by pQCD. 

In the context of QS, the present work  has shown that, thanks to its RG properties and  variational optimization prescription, the method can efficiently resum the perturbative series describing the pressure. 
Overall, this first application of the RGOPT EoS to NSQS gives remarkably good results when compared with the observational data suggesting that the method represents a very robust alternative to pQCD when dealing with the high density regime of QCD.
Possible extensions include the case $N_f=2+1$, the consideration of  the NNLO (three loop) contributions as well as the evaluation of a  hybrid EoS,  in order to describe the interior of neutron stars. 
\section*{Acknowledgments}
{T.E.R acknowledges support from Funda\c c\~ao Carlos Chagas Filho de Amparo \` a Pesquisa do Estado do Rio de Janeiro (FAPERJ), Process SEI-260003/002665/2021 and SEI-260003/019683/2022, and from Conselho
Nacional de Desenvolvimento Cient\'{\i}fico e Tecnol\'{o}gico (CNPq), Process No. 116037/2022-9. M.B.P. is partially supported by Conselho
Nacional de Desenvolvimento Cient\'{\i}fico e Tecnol\'{o}gico (CNPq),
Grant No  307261/2021-2  and by CAPES - Finance  Code  001.  T.E.R and M.B.P are also partially supported by
Instituto  Nacional  de  Ci\^encia  e Tecnologia de F\'{\i}sica
Nuclear e Aplica\c c\~{o}es  (INCT-FNA), Process No.  464898/2014-5. C.P is supported by  funds from FCT (Funda\c{c}\~ao para a Ci\^encia e a Tecnologia, I.P, Portugal) under the Projects No. UIDP/\-04564/\-2020 and 2022.06460.PTDC.}

\bibliographystyle{apsrev4-2}
\bibliography{bibliography}
\end{document}